\documentclass[12pt]{article}%
\usepackage{amsmath,latexsym}
\usepackage{graphicx}
\usepackage{amsmath}
\usepackage{amsfonts}
\usepackage{amssymb}%
\begin{document}

\title{{A simple argument for dark matter as an effect
    of slightly modified gravity}}
   \author{
  Peter K.F. Kuhfittig\\  \footnote{kuhfitti@msoe.edu}
 \small Department of Mathematics, Milwaukee School of
Engineering,\\
\small Milwaukee, Wisconsin 53202-3109, USA}

\date{}
 \maketitle

\begin{abstract}\noindent
This note presents a simple argument showing
that dark matter is an effect of $f(R)$ gravity
based on the definition of slightly modified
gravitational theories previously proposed
by the author.\\

\noindent
PAC numbers: 04.20.Jb, 04.50.Kd
\\
\\
Keywords: dark matter, $f(R)$ gravity
\end{abstract}

\section{Introduction}\label{E:introduction}

Discussions of dark matter and dark energy have
led to the hypothesis that Einstein's gravitational
theory does not hold on very large scales.  An
example of such a modification is the replacement
of the Ricci scalar $R$ by $f(R)$ in the
Einstein-Hilbert action to yield
\begin{equation}\label{E:f(R)}
   S_{f(R)}=\int\sqrt{-g}\,f(R)\,d^4x.
\end{equation}
Apart from the gravitational effects, no direct
evidence of dark matter has ever been found.  That
dark matter is a geometric effect of $f(R)$
gravity has already been shown by
B$\ddot{\text{o}}$hmer, Harko, and Lobo \cite
{BHL}.  All that is required is a small change in
the Ricci scalar.  The purpose of this note is to
present a simple alternative argument based on a
slight modification of $f(R)$ gravity as defined
by Kuhfittig \cite{pK13}.

\section{Galactic rotation curves}

An important objective in any modified
gravitational theory is to explain the peculiar
behavior of galactic rotation curves without
postulating the existence of dark matter: test
particles move with constant tangential velocity
$v_{\text{tg}}$ in a circular path.  It is noted
in Ref. \cite{BHL} that galactic rotation curves
generally show much more complicated dynamics.
For present purposes, however, the analysis can
be restricted to the region in which the velocity
is indeed constant.  In this note we will go a
step further and confine the analysis to a
narrow band around a constant $r=a$, the size
of which is to be specified later.  A constant
radius is of particular interest since for the
motion of particles in circular orbits, the
potential $V(r)$ must satisfy certain
conditions if the orbit is to be stable.

To meet the limited goals in this note, we
need only a few basic facts about galactic
rotation curves \cite{MGN, kN09, RKa, RKb,
RBFI}.  We start with the line element
\begin{equation}\label{E:line1}
ds^{2}=-e^{\phi(r)}dt^{2}+e^{\lambda(r)}dr^{2}
+r^{2}(d\theta^{2}+\text{sin}^{2}\theta\,d\phi^{2}),
\end{equation}
where $e^{\phi(r)}=B_0r^l$ and $l=2v^2_{\text{tg}}
\approx 0.000001$.  (We are using units in which
$c=G=1$.)  According to Ref. \cite{kN09}, for a
test particle with four-velocity $U^{\alpha}=
dx^{\alpha}/d\tau$ we have
\begin{equation}
   \dot{r}^2=E^2+V(r),
\end{equation}
where $\dot{r}=dr/d\tau$ and the potential is
\begin{equation}\label{E:V1}
   V(r)=-E^2+E^2\frac{e^{-\lambda}}{B_0r^l}-
   e^{-\lambda}\left(1+\frac{L^2}{r^2}\right).
\end{equation}
Moreover, if the circular orbits are given by
$r=a$, then
\[
    \frac{da}{d\tau}=0 \quad \text{and} \quad
    \left.\frac{dV}{dr}\right|_{r=a}=0.
\]
Also, two conserved quantities, the angular
momentum $L$ and the relativistic energy $E$, are
given by \cite{RKa, RKb}
\begin{equation}\label{E:L}
   L=\pm\sqrt{\frac{l}{2-l}}a
\end{equation}
and
\begin{equation}\label{E:E}
   E=\pm\sqrt{\frac{2B_0}{2-l}}a^{l/2}.
\end{equation}
The orbits are stable if
\begin{equation}\label{E:stable1}
   \left.\frac{d^2V}{dr^2}\right|_{r=a}<0.
\end{equation}

\section{Slightly modified $f(R)$ gravity}

As noted earlier, the radius $r=a$ will be
specified later.  For now it is sufficient to
recall that we will stay in a narrow band
around $r=a$.  The importance of this assumption
lies in the fact that, due to its simple form,
$e^{\phi(r)}=B_0r^l\approx\text{constant}$,
so we may assume that $\phi'\equiv 0$.
(By contrast, the behavior of $e^{\lambda(r)}$
is much more volatile in the present situation,
as we will see.)  The assumption
$\phi'\equiv 0$ allows us to define the
concept of ``slightly modified gravity,"
as discussed by Kuhfittig \cite{pK13}.
To that end, we list the gravitational field
equations in the form used by Lobo and
Oliveira \cite{LO09}:
\begin{equation}\label{E:Einstein1}
   \rho(r)=F(r)\frac{b'(r)}{r^2},
\end{equation}
\begin{equation}\label{E:Einstein2}
   p_r(r)=-F(r)\frac{b(r)}{r^3}
   +F'(r)\frac{rb'(r)-b(r)}{2r^2}
   -F''(r)\left[1-\frac{b(r)}{r}\right],
\end{equation}
\begin{equation}\label{E:Einstein3}
   p_t(r)=-\frac{F'(r)}{r}\left[1-\frac{b(r)}{r}
   \right]+\frac{F(r)}{2r^3}[b(r)-rb'(r)],
\end{equation}
where $F=\frac{df}{dR}$.  The curvature scalar is
given by
\begin{equation}\label{E:scalar}
  R(r)=\frac{2b'(r)}{r^2}.
\end{equation}
Observe that Eqs.
(\ref{E:Einstein1})-(\ref{E:Einstein3}) reduce
to the Einstein field equations for
$\phi'\equiv 0$ whenever $F\equiv 1$.
So comparing Eqs. (\ref{E:Einstein1}) and
(\ref{E:scalar}), a slight change in $F$ results
in a slight change in $R$.  According to Eq.
(\ref{E:f(R)}), this characterizes $f(R)$
modified gravity.  This change can be quantified
by assuming that $F(r)$ remains close to unity
and relatively flat, i.e., $|F'(r)|$ is relatively
close to zero.

\section{The solution}

Our first step is to expand $F(r)$ in a Taylor
series around $r=a$:
\[
  F(r)=F(a)+F'(a)(r-a)+\frac{1}{2}F''(a)(r-a)^2
  +\cdot\cdot\cdot.
\]
Since $r$ is assumed to be close to $r=a$,
higher-order terms become negligible.  So
\[
   F(r)=F(a)+F'(a)(r-a)
\]
and $F(r)|_{r=a}=F(a)\approx 1$, while
$F'(a)\approx 0$.  Observe that $F''$ can be
omitted.  Even though $r=a$ has not been specified,
we can assume that $F(a)$ and $F'(a)$ can be
treated as parameters in
Eqs. (\ref{E:Einstein1})-(\ref{E:Einstein3}) since
the precise values are not needed.

Next, we take the equation of state to be
\[
   p=m\rho, \quad 0<m<1,
\]
describing normal matter and also implying that
$p_r=p_t$, since a cosmological setting assumes a
homogeneous distribution of matter.  It now
follows from Eqs. (\ref{E:Einstein1}) and
(\ref{E:Einstein2}) that
\begin{equation*}
  mF\frac{b'}{r^2}=-F\frac{b}{r^3}+
  F'\frac{rb'-b}{2r^2}.
\end{equation*}
After some rearrangement, we get
\begin{equation*}
   \frac{b'}{b}=\frac{2F+rF'}{r(rF'-2mF)}=
   -\frac{1/m}{r}+\frac{2(1+m)F}{rF'-2mF}.
\end{equation*}
Integrating, we have (since $F$ and $F'$
are assumed to be constants)
\begin{equation*}
   \text{ln}\,b(r)=-\frac{1}{m}\text{ln}\,r
   +2(1+m)F\frac{1}{F'}\text{ln}
   |rF'-2mF|+\text{ln}\,c, \quad c>0,
\end{equation*}
or
\begin{equation}\label{E:shape1}
  b(r)=cr^{-1/m}|rF'-2mF|^{2(1+m)F/F'},
  \quad c>0.
\end{equation}
Thus
\begin{equation}\label{E:shape2}
  e^{-\lambda(r)}=1-\frac{b(r)}{r}=1-cr^{-1/m-1}
  |rF'-2mF|^{2(1+m)F/F'}, \quad c>0.
\end{equation}

Before returning to the stability question, we
need to show that thanks to the properties of
$F(r)$, $e^{-\lambda(r)}\approx\text{constant}$.
First recall that $|F'(a)|$ is close to zero
for any $a$.  So $e^{-\lambda(r)}$ in Eq.
(\ref{E:shape2}) could get large beyond any
bound as $F'(a)\rightarrow 0$.  (At this
point, $r=a$ is still unspecified.)  We
conclude that in order to get a meaningful
result, $|rF'(a)-2mF(a)|$ has to be less
than unity whenever $F'(a)>0$, thereby
causing $b(r)/r$ to become negligible for
$F'(a)$ sufficiently close to zero.   Since
$r$ is assumed to be close to $a$, by
letting $r\approx a$, we get the following
approximate inequality:
\[
  |aF'(a)-2mF(a)|\lesssim 1,\quad
      (r\approx a),
\]
or
\[
   -1\lesssim aF'(a)-2mF(a)\lesssim 1.
\]
Solving,
\begin{equation}\label{E:case1}
  \frac{2mF(a)-1}{F'(a)}\lesssim a
    \lesssim\frac{2mF(a)+1}{F'(a)},
\end{equation}
which is the approximate range of $a$
for the case $F'(a)>0$.  For example,
if $m\lesssim 1$ and $F\approx 1$, then
$1/F'(a)\lesssim a\lesssim 3/F'(a)$.
The interval widens to a large
region as $m$ gets closer to 1/2.  If
$F'(a)<0$, then we must have
\[
    |aF'(a)-2mF(a)|\gtrsim 1,
\]
leading to
\begin{equation}\label{E:case2}
   a\lesssim \frac{2mF(a)+1}{F'(a)}
   \quad \text{or} \quad
   a\gtrsim\frac{2mF(a)-1}{F'(a)},
\end{equation}
where the left inequality is disregarded since
$a$ is positive.  We now see that the previously
unspecified radius $a$ actually satisfies
the approximate inequalities (\ref{E:case1})
and (\ref{E:case2}), as a result of which
$e^{-\lambda(r)}\approx \text{constant}$.  (We
will also use these inequalities in the next
section to show that we do not obtain any
stable orbits in Einstein gravity.)

Returning now to Eq. (\ref{E:V1}), we can write
\begin{equation}\label{E:V2}
   V(r)=-E^2+e^{-\lambda}\left[\frac{2}{2-l}
   \left(\frac{a}{r}\right)^l-1-\frac{l}{2-l}
   \left(\frac{a}{r}\right)^2\right]
\end{equation}
and define
\[
   G(r)=\frac{2}{2-l}\left(\frac{a}{r}\right)^l
   -1-\frac{l}{2-l}\left(\frac{a}{r}\right)^2.
\]
Observe that $G(a)=0$ and
\[
  \left.
   G'(r)|_{r=a}=-\frac{a}{r^2}\left[\frac{2l}{2-l}
   \left(\frac{a}{r}\right)^{l-1}-\frac{2l}{2-l}
   \left(\frac{a}{r}\right)\right]\right|_{r=a}=0.
\]
Continuing with these calculations, we find that
\[
   G''(a)=-\frac{2l}{a^2}<0.
\]
So $G(r)<0$ in the neighborhood of $r=a$.
[See Fig. 1.]
\begin{figure}[tbp]
\begin{center}
\includegraphics[width=0.8\textwidth]{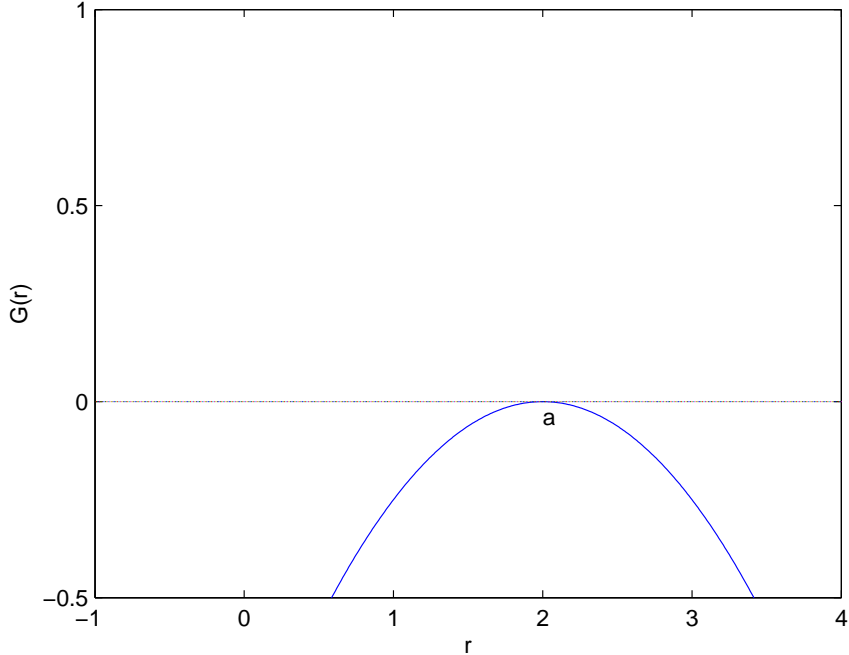}
\end{center}
\caption{$G(r)<0$ near $r=a$.}
\end{figure}

We now have $\dot{a}=0$ and $V'(a)=0$.  Since $-E^2$
and $e^{-\lambda}$ are constants, it follows
directly from Eq. (\ref{E:V2}) that
\begin{equation}\label{E:stable2}
   V''(a)<0,
\end{equation}
thereby showing that we have stable orbits.

\section{Comparison to Einstein gravity}

In this section we examine the limiting case
$F\rightarrow 1$, corresponding to Einstein
gravity.  Suppose in the equation of state
$p=m\rho$ we have $m\gtrsim \frac{1}{2}$.
Then inequality (\ref{E:case1}) yields
\begin{equation*}
   a\gtrsim\frac{2mF(a)-1}{F'(a)}>0.
\end{equation*}
But the same relationship is obtained from
inequality (\ref{E:case2}) whenever
$ m\lesssim \frac{1}{2}$.  Now, as we get
close to Einstein gravity, $F'(a)$ gets
close to zero regardless of the value of
$a$.  So in either case,
$a\rightarrow\infty$ as $F'\rightarrow 0$
and we do not get a (finite) stable orbit
in Einstein gravity.

.


\section{Conclusion}

This note gives a simple argument showing that
stable galactic orbits are an effect of $f(R)$
gravity without dark matter by assuming a
particular definition of slightly modified
gravity \cite{pK13}: the function $F(r)$ in the
Einstein field equations is characterized by the
properties $F(r)\approx 1$ and $|F'(r)|
\approx 0$.  Given the equation of state
$p=m\rho$, $0<m<1$, describing normal matter,
we obtain stable circular orbits, namely
$r=a$, satisfying the following approximate
inequalities: if $F'(a)>0$, then
\[
   \frac{2mF(a)-1}{F'(a)} \lesssim a
   \lesssim\frac{2mF(a)+1}{F'(a)};
\]
if $F'(a)<0$, then
\[
  a\gtrsim\frac{2mF(a)-1}{F'(a)}.
\]
These inequalities indicate that such stable orbits
exist in a large region.  There are no stable orbits
in the limiting case, $F\rightarrow 1$,
$F'\rightarrow 0$, corresponding to Einstein
gravity, without hypothesizing the existence
of dark matter.

\end{document}